\documentclass[aps,prl,twocolumn,groupedaddress,floatfix,superscriptaddress]{revtex4}
\usepackage{epsfig}
\usepackage[dvipsnames,usenames]{color}
\usepackage{enumerate}
\usepackage{color}
\usepackage{colordvi}
\usepackage{amsmath}
\usepackage{amssymb}
\usepackage{comment}
\usepackage{graphicx}

\emergencystretch=\maxdimen
\begin{document}

\title{Cooling Atomic Gases With Disorder}

\author{Thereza Paiva}
\affiliation{Departamento de F\'\i sica dos S\'olidos, Instituto de F\'\i sica, Universidade Federal do Rio de Janeiro, 21945-970, Rio de Janeiro, RJ, Brazil}
\author{Ehsan Khatami}
\affiliation{Department of Physics, San Jose State University, San Jose, CA 95616, USA}
\author{Shuxiang Yang}
\affiliation{Department of Physics and Astronomy, Louisiana State University, Baton Rouge, LA 70803, USA}
\author{Valery Rousseau}
\affiliation{Department of Physics and Astronomy, Louisiana State University, Baton Rouge, LA 70803, USA}
\author{Mark Jarrell}
\affiliation{Department of Physics and Astronomy, Louisiana State University, Baton Rouge, LA 70803, USA}
\author{Juana Moreno}
\affiliation{Department of Physics and Astronomy, Louisiana State University, Baton Rouge, LA 70803, USA}
\author{Randall G. Hulet}
\affiliation{Department of Physics and Astronomy and Rice Quantum Institute, Rice University, Houston, TX 77005, USA}
\author{Richard T. Scalettar}
\affiliation{Department of Physics, University of California, Davis, CA 95616, USA}

\begin{abstract}
Cold atomic gases have proven capable of emulating a number of
fundamental condensed matter phenomena including Bose-Einstein
condensation, the Mott transition, Fulde-Ferrell-Larkin-Ovchinnikov 
pairing and the quantum Hall effect.  Cooling to a low enough
temperature to explore magnetism and exotic superconductivity
in lattices of fermionic atoms remains a challenge. We propose a method
to produce a low temperature gas by preparing it in a disordered potential
and following a constant entropy trajectory to deliver the
gas into a non-disordered state which exhibits these incompletely
understood phases.
We show, using quantum Monte Carlo simulations, that 
we can approach  the Ne\'el temperature of the three-dimensional Hubbard model
 for experimentally achievable parameters.  Recent
experimental estimates suggest the randomness required lies
in a regime where atom transport and equilibration are still robust.
\end{abstract}

\maketitle

%%%%%%%%%%%%%%%%%%%%%%%%%%%%%%%%%%%%%%%%%%%%%%%%%%%%%%%%%%%%%%%%%%%%%%%%%%%%%%%%
% INTRODUCTION
%%%%%%%%%%%%%%%%%%%%%%%%%%%%%%%%%%%%%%%%%%%%%%%%%%%%%%%%%%%%%%%%%%%%%%%%%%%%%%%%
\vskip0.10in \noindent
{\bf Introduction:}  The interplay of disorder and interactions is a
central problem in condensed matter physics, both from the viewpoint of
materials like the heavy fermions~\cite{maclaughlin96,curro09},
high-temperature superconductors~\cite{keimer15}, and
manganites~\cite{dagotto05}, and also because of intriguing theoretical
issues such as the fate of Anderson localization in the presence of
interactions, especially in two dimensions~\cite{lee85,kravchenko94}.
Ultracold atomic gases offer the opportunity to emulate these
fundamental issues using optical speckle~\cite{Billy08, pasienski10},
impurities \cite{Gadway11}, or a quasiperiodic optical lattice
\cite{Roati08,Fallani07} to introduce randomness.  In the bosonic case,
the competition between strong interactions and strong disorder has been
studied in the context of the elusive Bose glass phase \cite{Fallani07,
pasienski10, Gadway11}, while for fermions, a recent experiment has
explored disorder-induced localization in the three-dimensional (3D)
Hubbard model of strongly-interacting fermions \cite{kondov15}.

In this paper, we explore the thermodynamics of interacting, disordered
systems and suggest that, in addition to studies of the many-body phenomena noted
above, preparing a gas in a random potential 
might be exploited to cool the atoms. 
Specifically, we show using an unbiased numerical method
that 
one can lower the temperature and access the regime with 
long-range magnetic order by adiabatically decreasing the randomness in the chemical
potential or hopping energies of the Hubbard Hamiltonian.
%the temperature $T$ decreases along
%adiabatic curves of the Hubbard Hamiltonian as randomness in the chemical
%potential or hopping energies is decreased.
Results for the double occupancy and antiferromagnetic
structure factor lend physical insight into this effect. We also
present arguments, partially based on recent experiments, that our
suggestion is achievable in practice.

We consider the disordered Hubbard
Hamiltonian,
\begin{align}
H&=- \sum_{\langle ij \rangle \sigma}
t^{\phantom{\dagger}}_{ij} 
( c^{\dagger}_{i\sigma} c^{\phantom{\dagger}}_{j\sigma} +
 c^{\dagger}_{j\sigma} c^{\phantom{\dagger}}_{i\sigma} )
\nonumber \\
&+ U \sum_i (n^{\phantom{\dagger}}_{i\uparrow}-\frac{1}{2})
( n^{\phantom{\dagger}}_{i\downarrow}-\frac{1}{2})
- \sum_i \mu^{\phantom{\dagger}}_i \, 
( n^{\phantom{\dagger}}_{i\uparrow} 
+ n^{\phantom{\dagger}}_{i\downarrow} )
\end{align}
whose emulation~\cite{jaksch98,esslinger10} with optical lattices is
possible using two hyperfine species of fermionic atoms.  Here
$c^{\dagger}_{i \sigma} (c^{\phantom{\dagger}}_{i \sigma})$ is the
creation (destruction) operator for a fermion at spatial site $i$ and
spin (or hyperfine state) $\sigma$.  We consider a cubic lattice of $N$
sites, and hopping $t_{ij}$ between near neighbors $\langle ij \rangle$.
The hopping, and the on-site repulsion $U$, can be tuned with the
lattice depth and the Feshbach
resonance~\cite{jaksch98}, allowing for the successful exploration of the
Mott transition~\cite{jordens2008,schneider2008,p_duarte_15}.

Disorder is introduced via a spatially random chemical potential $\mu_i$
or hopping $t_{ij}$.  We choose uniform distributions $\mu_0 -\Delta_\mu
< \mu_i < \mu_0+\Delta_\mu$ or $t_0-\Delta_t < t_{ij} < t_0+\Delta_t$,
and set the mean of the hopping energy $t_0=1$ as the energy scale. For
most of this paper we choose $\mu_0=0$, which makes the lattice
half-filled (average density $n=1$). However, we also gain insight into the
effects of a confining potential, in which the chemical potential
increases as one moves spatially away from the trap
center, by presenting data for different densities.

%%%%%%%%%%%%%%%%%%%%%%%%%%%%%%%%%%%%%%%%%%%%%%%%%%%%%%%%%%%%%%%%%%%%%%%%%%%%%%%%
% FIGURE 1
%%%%%%%%%%%%%%%%%%%%%%%%%%%%%%%%%%%%%%%%%%%%%%%%%%%%%%%%%%%%%%%%%%%%%%%%%%%%%%%%
\begin{figure}[t]
\includegraphics[width=0.9\columnwidth]{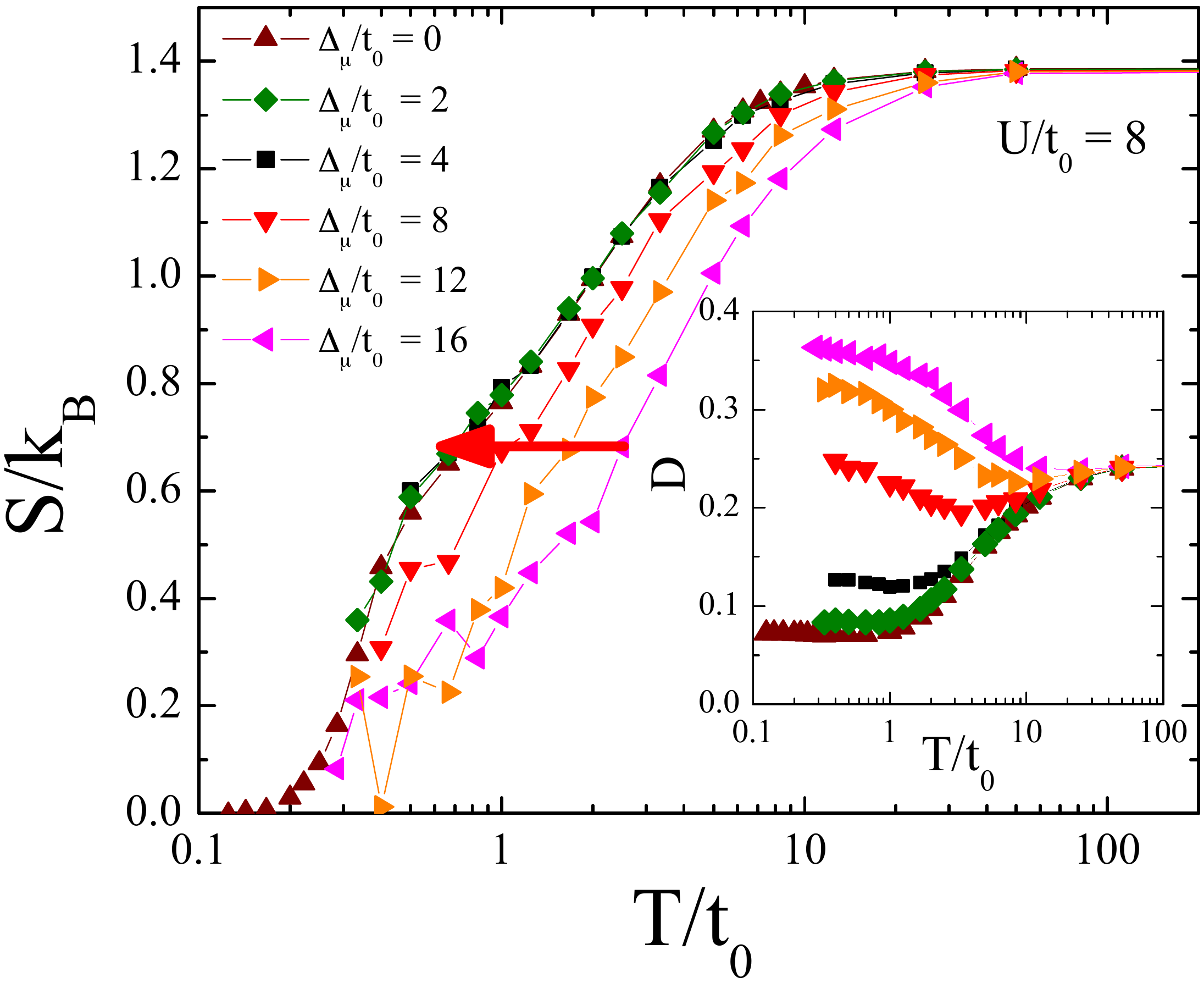}
\caption{
Entropy $S$ as a function of temperature $T$ for
different site disorder strengths $\Delta_\mu$ at $U/t_0=8$.  
%\textcolor{red}{In the absence of disorder, $S(T)$ has large slope
%$dS/dT = C/T$ at a higher temperature, where moment formation occurs,
%and at a lower temperature, where moment ordering occurs.
$S$ is largely independent of disorder strength for $\Delta_\mu/t_0 =2,4
\lesssim U/t_0 = 8$.  For larger randomness,
$S(T)$ decreases with $\Delta_\mu$ so that if disorder
is turned off adiabatically, the temperature $T$ decreases, as indicated
by the horizontal arrow at $S/k_B={\rm ln}\,2$.  The inset shows the
double occupancy $D(T)$.  Large disorder $\Delta_\mu$ changes the sign
of the slope $dD/dT$ from mostly positive, to mostly negative.  Here, and
in all subsequent figures, unless otherwise indicated, the lattice size
is $6^3$, the density $n=1$ is at half filling, and the Trotter
discretization is $\Delta \tau = 1/(20t_0)$. Up to 300 disorder realizations 
are used in the disorder averages.
\label{fig:SU8}
}
\end{figure}

Our computational method,  determinant quantum Monte Carlo (QMC)
\cite{blankenbecler1981,white1989}, treats disorders and interactions on
an equal, exact footing, and provides a solution to the Hubbard Hamiltonian on 
lattices of finite spatial size.  We focus on the disorder dependence of the 
entropy $S(T)$, obtained via a thermodynamic integration of the
energy~\cite{dare07} down from $T=\infty$.  We also report results for
the (site-averaged) double occupancy $D=1/N \sum_i \langle n_{i\uparrow}
n_{i\downarrow} \rangle$, and the structure factor $S_{\bf q}=\sum_r e^{i{\bf q}\cdot {\bf r}}c({\bf r})$ at
${\bf q}=(\pi,\pi,\pi)$; $S_\pi$, where $c({\bf r}) = \langle c^{\dagger}_{i+{\bf r} \,
\downarrow} c^{\phantom{\dagger}}_{i+{\bf r} \, \uparrow} c^{\dagger}_{i \,
\uparrow} c^{\phantom{\dagger}}_{i \, \downarrow} \rangle$ are spin-spin 
correlation functions.

%%%%%%%%%%%%%%%%%%%%%%%%%%%%%%%%%%%%%%%%%%%%%%%%%%%%%%%%%%%%%%%%%%%%%%%%%%%%%%%%
% RESULTS
%%%%%%%%%%%%%%%%%%%%%%%%%%%%%%%%%%%%%%%%%%%%%%%%%%%%%%%%%%%%%%%%%%%%%%%%%%%%%%%%
\vskip0.10in \noindent
{\bf Results:} The effect of site disorder on $S(T)$ is shown in
Fig.~\ref{fig:SU8} at $U/t_0=8$, where the Ne\'el transition temperature
($T_N$) in the homogeneous 3D Hubbard model attains its maximal
value~\cite{staudt2000}.  $S(T)$ is largely unaffected by disordered
site energies until $\Delta_\mu$ becomes comparable to $U$.  This is a
consequence of the fact that for temperatures  less
than the repulsion $U$, the Hubbard model has the character of a Mott
insulator in which $U$ blocks transport of Fermions away from singly
occupied sites.  Such a Mott state is immune to the effects of small
disorder $\Delta_\mu/U \lesssim 1$.  Our calculated entropy $S(T)$ and
double occupancy $D(T)$ (see the inset of Fig.~\ref{fig:SU8}) confirm this picture.
However, when $\Delta_\mu/U \gtrsim 1$ the entropy curves shift
systematically to higher $T$, reflecting a disorder-driven decrease in
$S$ at constant $T$. The reduction in $S$ can be viewed as the transfer
of weight in the specific heat $C(T)$ to a higher temperature:
Disorder suppressing the peak in $C(T)$ associated with
local magnetic ordering at the exchange energy scale $T \sim J =
4t_0^2/U$, and increasing $C(T)$ at a higher $T$ that scales like $\Delta_{\mu}$ due to
excitations arising from the transfer of charge between sites of
different local $\mu_i$~\cite{SM}. It is expected
 that at very low temperatures, the disorder increases the degeneracy of
 the low lying states, and hence the entropy.  However, our results
 indicate that in the temperature range of interest, $T \gtrsim T_N$ (the Ne\'{e}l temperature), 
 disorder reduces $S$.

%%%%%%%%%%%%%%%%%%%%%%%%%%%%%%%%%%%%%%%%%%%%%%%%%%%%%%%%%%%%%%%%%%%%%%%%%%%%%%%%
% FIGURE 2
%%%%%%%%%%%%%%%%%%%%%%%%%%%%%%%%%%%%%%%%%%%%%%%%%%%%%%%%%%%%%%%%%%%%%%%%%%%%%%%%
\begin{figure}[t]
\includegraphics[width=0.9\columnwidth]{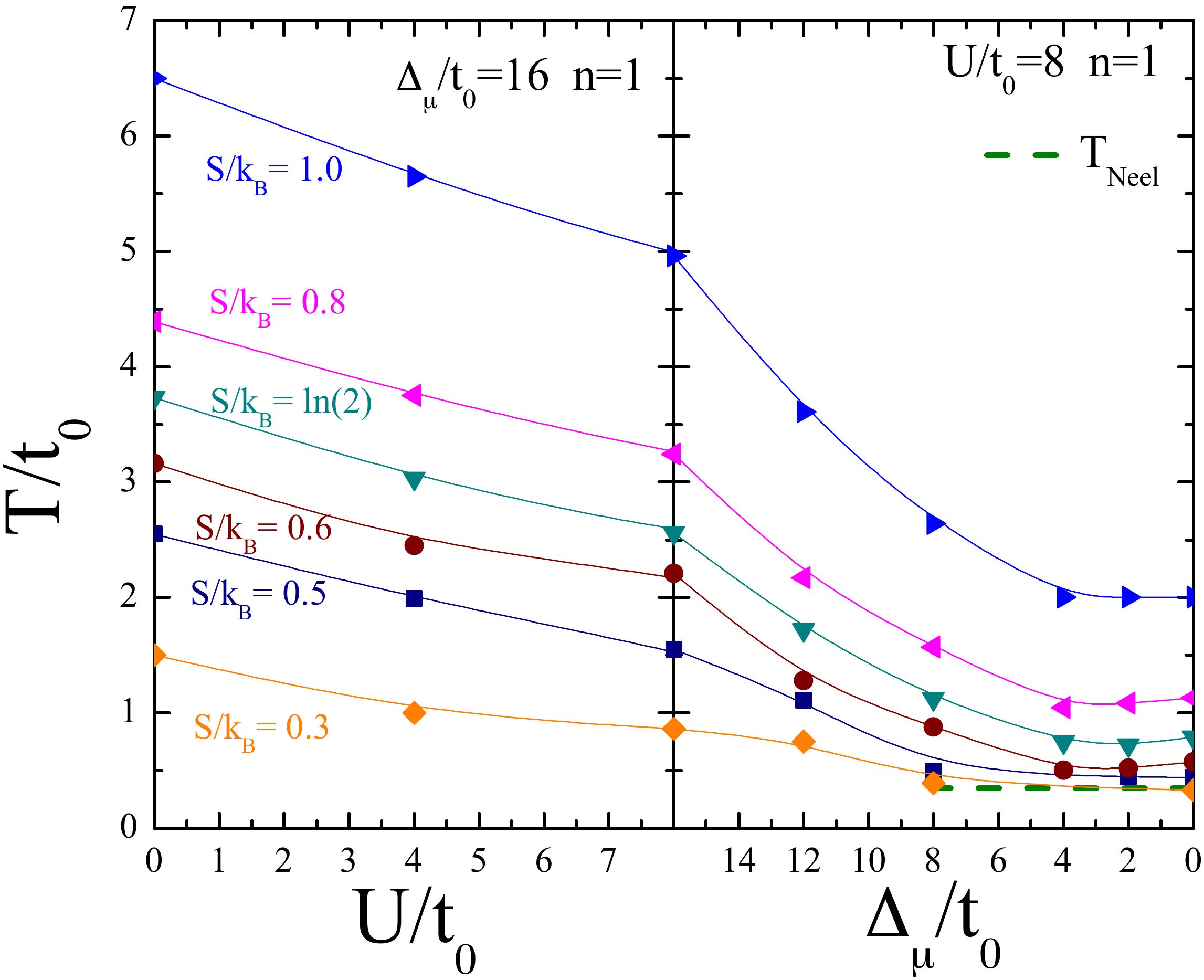}
\caption{
Adiabats of the disordered 3D Hubbard model along a path which combines
an increase of the interaction strength from $U/t_0=0$ to $U/t_0=8$ at
fixed $\Delta_\mu/t_0=16$ followed by a reduction of the site disorder.
$T/t_0$ decreases along both trajectories, and, in particular, by about
a factor of three at fixed $S/k_{\rm B}={\rm ln}\,2$ along the second
path.  For $S/k_{\rm B}=0.5$, the same reduction brings $T$ down to near
$T_N$. 
\label{fig:TvsDelta}
}
\end{figure}

The family of $S(T)$ curves in Fig.~{\ref{fig:SU8} indicates that if
$\Delta_\mu$ is switched to zero at constant entropy, the temperature
$T$ decreases, in analogy to Pomeranchuk cooling which occurs 
in a non-disordered lattice when the
ratio of repulsion to hopping $U/t_0$ is increased
adiabatically~\cite{werner2005}.  For the case of site disorder, the double occupancy
shows a negative slope $dD/dT<0$ as seen in the inset of  Fig.~{\ref{fig:SU8}.  At high enough temperatures $T
\gtrsim t_0,U,\Delta_\mu$, up and down spin fermions are uncorrelated,
and $D$ factorizes, $D = \langle n_{i \uparrow} n_{i\downarrow} \rangle
\rightarrow \langle n_{i \uparrow} \rangle \langle n_{i \downarrow}
\rangle$ ($=1/4$ at half-filling).  In the clean limit, as $T$ is
lowered, the on-site repulsion eliminates double occupancy, and $D$
falls. Disordered site energies reduce the penalty for double occupancy
from $U$ to $U_{\rm eff} = U 
- |\mu_i - \mu_j|$ so that as $\Delta_\mu$ grows, $U_{\rm eff}$ becomes
  negative.  The low $T$ phase consists predominantly of doubly 
occupied and empty sites so that in the limit $\Delta_\mu/t_0 \gg 1$,  $D$
approaches $\frac{1}{2}$.

From Fig.~\ref{fig:SU8} we can infer the behavior of $T$ as
$\Delta_\mu/t_0$ is lowered adiabatically at {\it fixed} $U/t_0$.
Optical lattice experiments, however, typically involve an increase of
$U/t_0$ from zero to its final value. Figure \ref{fig:TvsDelta} presents
the adiabatic curves of a combined protocol in which the
interaction is increased from $U/t_0=0$ to $U/t_0=8$ in
the presence of fixed disorder $\Delta_\mu/t_0=16$, followed by the
suppression of the disorder to $\Delta_\mu/t_0=0$. Data are shown for
different values of the starting entropy $S/k_B$. Figure
\ref{fig:TvsDelta} contains the central observation of our paper: a
significant decrease in temperature results from following these
adiabats.  The substantial cooling in the second part of the path, at
fixed $U/t_0$, is implicit in Fig.~\ref{fig:SU8}.  A reduction in
$T/t_0$ also occurs in the initial turning on of the interaction, more
so in the presence of disorder than occurs in the clean system
\cite{fuchs11,paiva11}. Our QMC results indicate that beginning at
temperatures $T/t_0 \lesssim  2.5$ at $\Delta_\mu = 16t_0$ would be
sufficient to reach $T_{N}$ by the time the clean limit is reached.
%%{\color{blue} 
However, an important question arises: Can
the trapped system in the presence of disorder be cooled 
down to an initial temperature $T/t_0\sim1.5$, or possibly even lower,  
close to what is initially needed for the clean system to reach the Ne\'{e}l 
phase ($T_N/t_0\sim 0.35$)~\cite{paiva11}.
Current cooling capabilities have achieved a final temperature of $T/t_0=0.5$
($1.4 T_N$) for $U/t_0 \sim 11$ at the trap center~\cite{hart15}. We provide several 
suggestions concerning its feasibility in our concluding remarks.
%} A critical question
%remains: can the trapped disordered system be initially cooled to the 
%same, or lower, temperatures as the clean system? We will discuss 
%different possibilities in the following sections.

Since random $\mu_i$ and $t_{ij}$ occur together with optical
speckles~\cite{white09,zhou10}, we also explore the case of bond
disorder.  Figure \ref{fig:SvsTdt} shows $S(T)$ for nonzero $\Delta_t$
(and $\Delta_\mu=0$).  Significant disorder-induced cooling occurs.   It
is notable that $\Delta_t/t_0 \sim 1$ is sufficient to produce an effect
on the entropy, whereas the scale of random site energies required to
change $S$ is much larger (Fig.~\ref{fig:SU8}).  
This is a consequence of the fact that random hopping immediately leads to
a range of exchange energies $J_{ij} \sim 4t_{ij}^2/U$ which reduces
the moment ordering.  Random $\mu_i$ also smear $J_{ij}$
but, since they are added to $U$ in the energy denominator, initially
have only a small effect. Random hopping thus offers
cooling at lower temperature (entropy) scales for $\Delta_t \sim t_0$
than does random chemical potential, without requiring a `threshold value',
$\Delta_\mu>U$. Unlike for the chemical potential disorder,
the basic structure of $D(T)$ remains unaltered for the clean
system~\cite{SM}.

%%%%%%%%%%%%%%%%%%%%%%%%%%%%%%%%%%%%%%%%%%%%%%%%%%%%%%%%%%%%%%%%%%%%%%%%%%%%%%%%
% FIGURE 3
%%%%%%%%%%%%%%%%%%%%%%%%%%%%%%%%%%%%%%%%%%%%%%%%%%%%%%%%%%%%%%%%%%%%%%%%%%%%%%%%
\begin{figure}[t]
\includegraphics[width=0.9\columnwidth]{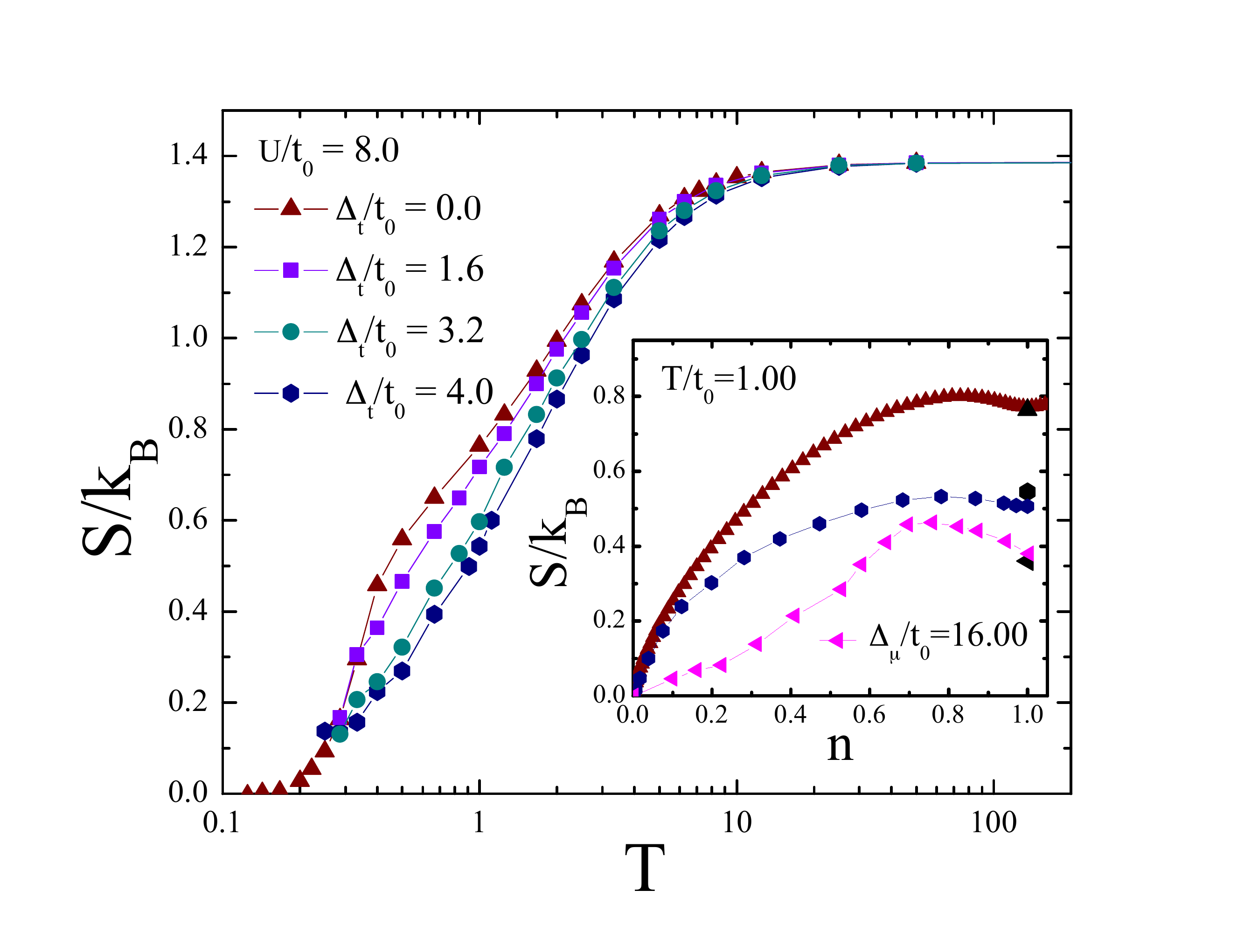}
\caption{Entropy versus temperature for hopping disorder. Here, disorder
cooling is strongest at lower entropies $S \sim 0.5$. The inset shows
the entropy as a function of density of the clean system for
$\Delta_{\mu}/t_0=16$ and $\Delta_t/t_0=4$ at fixed $U/t_0=8$ and
$T/t_0=1$. Here, the entropy is obtained using
$S(\mu,T)=\int_{-\infty}^{\mu}d\mu \frac{\partial n}{\partial T}|_{\mu}$
\cite{Val}, except for the three data points in black (darker shade) at
$n=1$, which are obtained via integrating over $\beta$.
\label{fig:SvsTdt}
}
\end{figure}

To provide some insight into possible effects of the inhomogeneous
densities resulting from a confining potential, we show the entropy
as a function of density for the clean system and for chemical
potential disorder $\Delta_\mu/t_0=16$ and hopping disorder
$\Delta_t/t_0=4$ in the inset of Fig.~\ref{fig:SvsTdt}.  Although there
is some structure to the curves, entropy is systematically lowered for
all densities as disorder is introduced. Thus disorder cooling is not a
special feature of half-filling, but likely occurs for a broad range of
densities.

We note that there are important questions of principle which would
arise in a full treatment of a trap~\cite{paiva11,fuchs11}. QMC 
calculations for clean systems employed a set of homogeneous simulations, combined with
the local density approximation (LDA), to understand how the density, 
double occupancy, and entropy are inhomogeneously distributed in a system 
with smoothly varying chemical potential. This is a considerably harder 
task in the presence of disorder, because the implementation, and indeed 
even the validity, of the LDA is much less straightforward with a rapidly 
varying $\mu_i$ or $t_{ij}$.  In fact, the LDA has the curious
feature that thermodynamic properties are insensitive to the specific
geometric organization of the sites with the different chemical
potentials:  The local entropy $s_{\mu_i}$ is unaltered for any two
systems with the same collection $\{\mu_i\}$ whether they are randomly
distributed or ordered spatially in some pattern, a patently unphysical
result.

%%%%%%%%%%%%%%%%%%%%%%%%%%%%%%%%%%%%%%%%%%%%%%%%%%%%%%%%%%%%%%%%%%%%%%%%%%%%%%%%
% FURTHER ANALYSIS
%%%%%%%%%%%%%%%%%%%%%%%%%%%%%%%%%%%%%%%%%%%%%%%%%%%%%%%%%%%%%%%%%%%%%%%%%%%%%%%%
\vskip0.10in \noindent
{\bf Further Analysis:}   
Observing the onset of long-range antiferromagnetic (AF) correlations
is a central goal of the field. To see the development of these correlations
as the disorder is turned off, we show in Fig.
\ref{fig:Spi} the structure factor $S_\pi$
as a function of $T$ for different site (top panel) and bond (bottom
panel) disorder strengths. $\Delta_\mu> U$ completely destroys the sharp
rise in $S_\pi$, which occurs here on a $6^3$ lattice at a value close
to the bulk $T_N/t_0 \sim 0.35$ for $U/t_0=8$.  The suppression of
magnetic order is a consequence of the destruction of the local moments
$m^2 = \left < (n_{i\uparrow} - n_{i\downarrow})^2\right > = 1 - 2D$ at
half-filling (see the inset of Fig.~\ref{fig:SU8}).  $S_\pi $ is also
suppressed by $\Delta_t$ despite the fact that it has only a small
effect on $m^2$~\cite{SM}. The likely mechanism for the destruction of
AF order in this case is the introduction of fluctuations in the
near-neighbor exchange $J_{ij} \sim 4 t_{ij}^2/U$. As a consequence, of
this anisotropy, singlets can form on the bonds with large $J_{ij}$.
When many pairs of sites are effectively removed from the lattice, order
is lost. Although both bond and site disorder reduce $S_\pi$, it is
important to emphasize that low $T$ is reached by turning the disorder
{\it off}, so that the terminal state is the sought after regime of
large AF correlations.

%%%%%%%%%%%%%%%%%%%%%%%%%%%%%%%%%%%%%%%%%%%%%%%%%%%%%%%%%%%%%%%%%%%%%%%%%%%%%%%%
% FIGURE 4
%%%%%%%%%%%%%%%%%%%%%%%%%%%%%%%%%%%%%%%%%%%%%%%%%%%%%%%%%%%%%%%%%%%%%%%%%%%%%%%%
\begin{figure}[t]
\includegraphics[width=0.9\columnwidth]{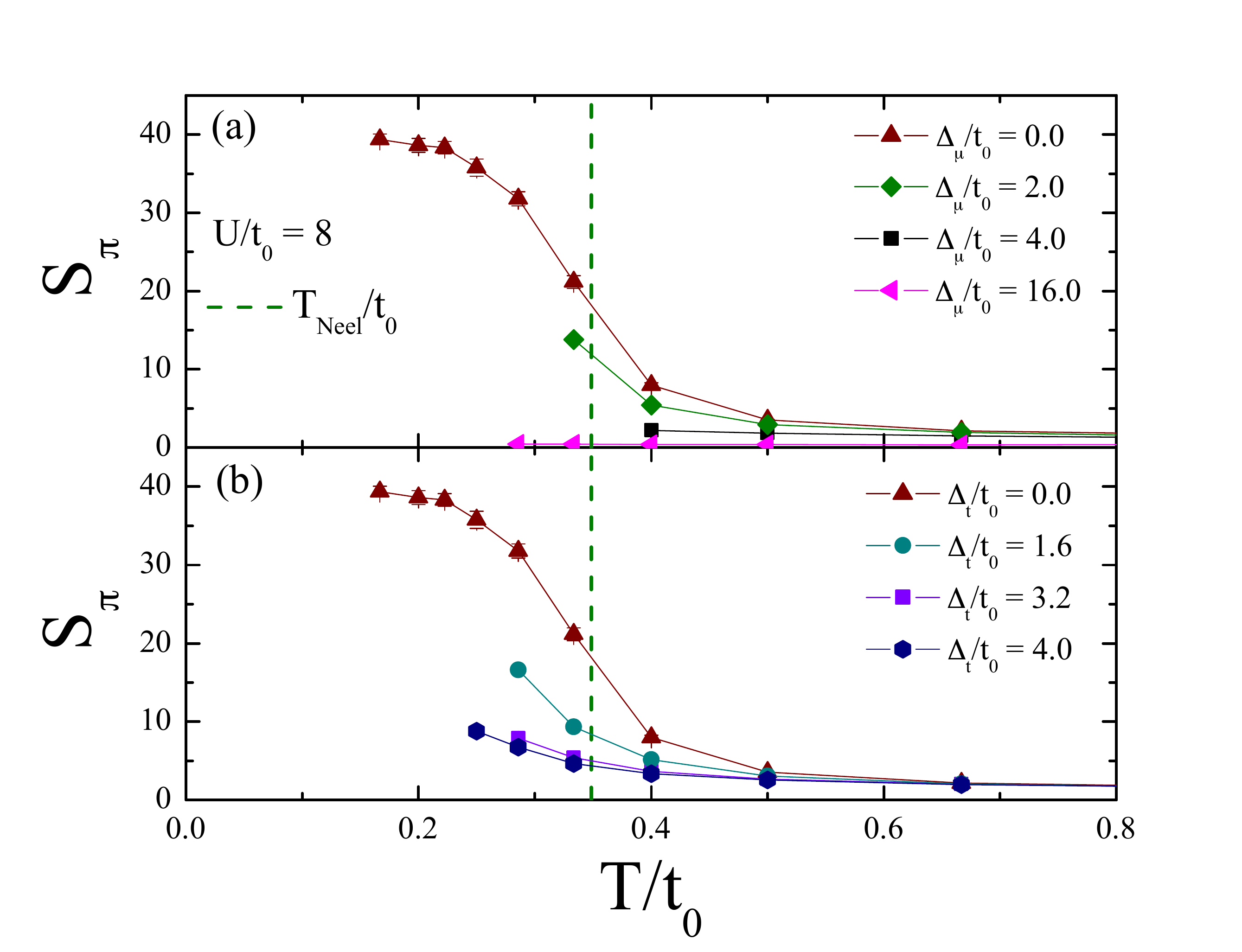}
\caption{
Antiferromagnetic structure factor $S_\pi$ as a function of $T$ as the
site (top) and bond (bottom) disorder is varied.  Site disorder drives
$S_\pi$ to zero for $\Delta_\mu \gtrsim U$ by destroying the magnetic
moments $m^2=1-2D$, whereas singlet formation on bonds with large
$J_{ij} \sim t_{ij}^2/U$ is induced by sufficiently large $\Delta_t$ and
also destroys the AF long-range order.
\label{fig:Spi}
}
\end{figure}

Equilibration is crucial to the viability of disorder cooling. Recent
experiments by the DeMarco group~\cite{kondov15} provide evidence that
the requisite $\Delta_\mu$ lie well below the threshold where randomness
drives atomic velocities to zero:  Measurements of mass transport show
that the center-of-mass velocity only vanishes above $\Delta_\mu/t_0
\sim 21.7 \pm 1.6$ for $U/t_0=3.8$ and $\Delta_\mu/t_0 \sim 31.7 \pm
4.2$ for $U/t_0=9.1$. The implications of these results for disorder
cooling are considered in Fig.~\ref{fig:TfTi}, which shows the final
temperature $T_f(T_i,\Delta_\mu)$ which would result from starting at
initial temperature $T_i$ and disorder $\Delta_\mu$, and turning off
randomness adiabatically.  Figure \ref{fig:TfTi} complements
Fig.~\ref{fig:TvsDelta} and provides another way of analyzing the
lowering of $T_f$ starting from states at $T_i$ with $\Delta_\mu$ beyond
$U/t_0$ and adiabatically following a path to $\Delta_\mu=0$. The
reduction in temperature, $T_i-T_f$, can be as large as $0.65t_0$ for
$\Delta_\mu/t_0=16$ and $U/t_0=4$, starting at $T_i/t_0=1$ and $1.35t_0$
for $T_i/t_0=2$. The many-body localization (MBL) critical disorder
strengths for $U/t_0= 4.0, 8.0$ and 12.0 (denoted by horizontal arrows
in Fig.~\ref{fig:TfTi}) lie above the range which provides substantial
cooling.  These comparisons provide considerable support to the
likelihood that equilibration will still occur in the regime where
disorder-induced cooling is effective.

%%%%%%%%%%%%%%%%%%%%%%%%%%%%%%%%%%%%%%%%%%%%%%%%%%%%%%%%%%%%%%%%%%%%%%%%%%%%%%%%
% FIGURE 5
%%%%%%%%%%%%%%%%%%%%%%%%%%%%%%%%%%%%%%%%%%%%%%%%%%%%%%%%%%%%%%%%%%%%%%%%%%%%%%%%
\begin{figure}[t]
\includegraphics[width=0.9\columnwidth]{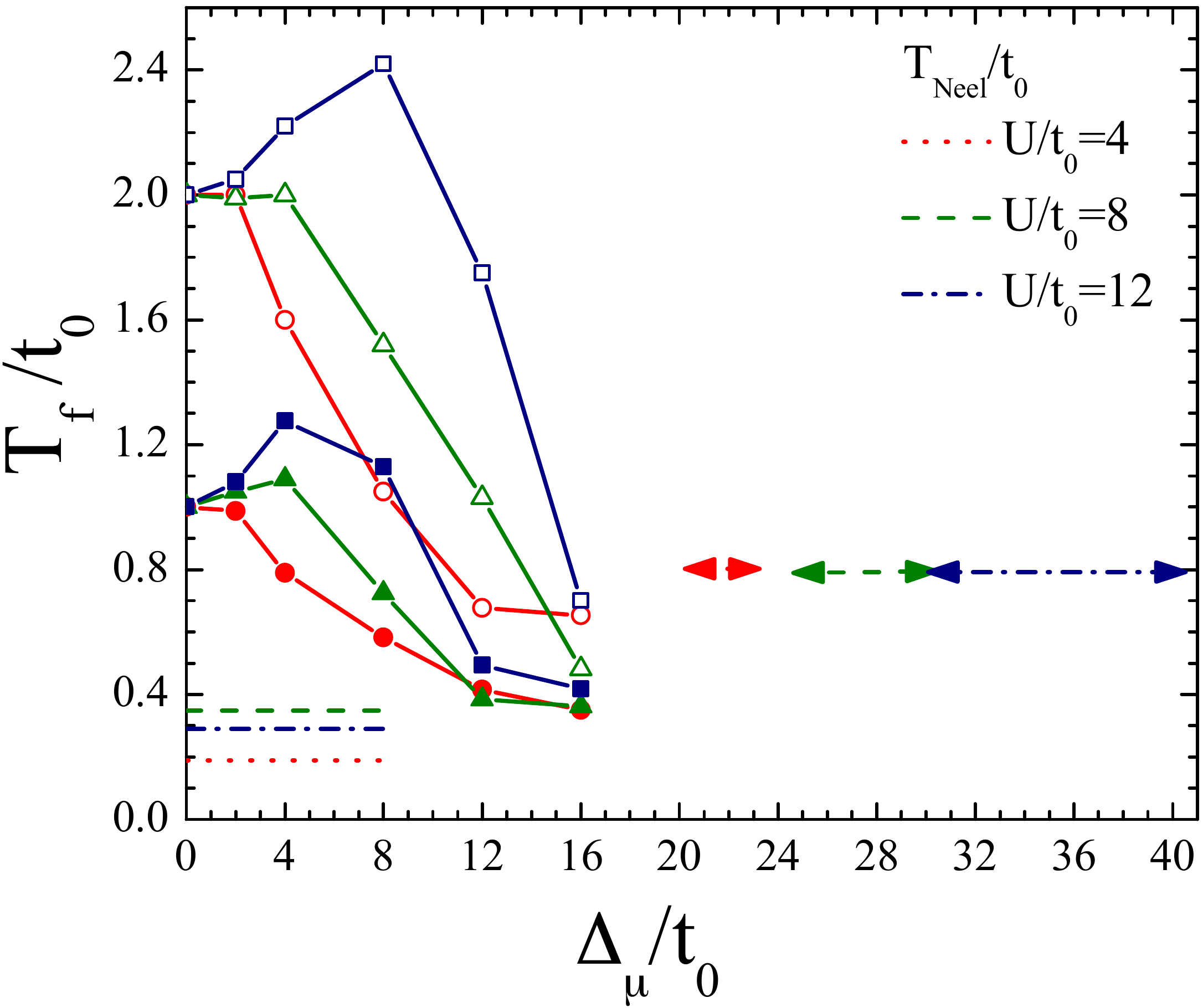}
\caption{
Final temperature $T_f/t_0$ resulting from adiabatically turning off disorder 
in a system with initial disorder strength $\Delta_\mu/t_0$, shown 
on the horizontal axis, and two different initial temperatures $T_i/t_0=1,\ 2$.
As $\Delta_\mu$ increases beyond $U/t_0$, $T_f$ decreases.  The Ne\'el
temperatures $T_N/t_0=0.19, 0.35, 0.29$ for $U/t_0=4,8,12$, respectively, are 
shown as dashed horizontal lines. The horizontal arrows are estimates for the 
onsets of MBL for (from left to right) $U/t_0=4.0, 8.0, 12.0$, obtained by 
linearly interpolating $\Delta_c/12t_0$ vs $U/12t_0$ in Fig.~3 of Ref.
\cite{kondov15}.  
\label{fig:TfTi}}
\end{figure}

{\bf Implementation and Concluding Remarks:} The feasibility of
disordered-induced cooling depends on the ability to realize low initial
temperatures in the disordered lattice. Since turning on disorder heats
the gas, this energy must be removed before attempting to cool more
deeply using our method. There have been no direct attempts to cool in a
disordered lattice, but several schemes are promising.  One such method
is sympathetic cooling by another atomic species \cite{LeBlanc07} or
spin-state of the same species \cite{Deutsch98,Jaksch99} that by proper
choice of lattice wavelength or polarization is unaffected by the
lattice.   Another approach is to implement a compensated lattice, where
the overall confinement created by the infrared lattice beams is
compensated by overlapping blue-detuned beams \cite{Mathy12}.  By tuning
the intensity of the blue-detuned beams the threshold for evaporation
can be brought near the chemical potential, resulting in very low
temperatures \cite{hart15}.  While this scheme has only been implemented
in a clean lattice, it seems plausible that it can work in any situation
where there is sufficient mobility.

A second approach is to mask the disorder in such a way that it is
applied only to a small spatial subregion of the entire gas. Through
thermal contact, atoms in this region could be cooled by the larger
reservoir region outside the disordered volume.  If the clean gas is
then discarded, one again has the starting point of a disordered gas at
the same initial $T$ as a clean one.  Complex optical potentials to
perform these roles can be created using phase-imprinting spatial light
modulators \cite{McGloin03,Gaunt13} or micro-mirror devices
\cite{Brandt11}.

%%%%%%%%%%%%%%%%%%%%%%%%%%%%%%%%%%%%%%%%%%%%%%%%%%%%%%%%%%%%%%%%%%%%%%%%%%%%%%%%
% ACKNOWLEDGEMENTS
%%%%%%%%%%%%%%%%%%%%%%%%%%%%%%%%%%%%%%%%%%%%%%%%%%%%%%%%%%%%%%%%%%%%%%%%%%%%%%%%
\acknowledgements

T.P. and R.T.S. thank the CNPq Science Without Borders program. V.R.,
J.M. and M.J. acknowledges support from the National Science Foundation
(NSF) Grant No.  OISE-0952300. The work at Rice was supported by the
NSF, the Welch Foundation (Grant No. C-1133), and ARO-MURI Grant No.
W911NF-14-1-0003. Insightful conversations with Dan Stamper-Kurn and
Brian DeMarco are gratefully acknowledged.

\end{document}

\title{Supplemental Material:\\
Cooling Atomic Gases With Disorder}

\author{Thereza C. L. Paiva}
\affiliation{Departamento de Fisica dos S\'olidos, Instituto de F\'isica, Universidade Federal do Rio de Janeiro, Cx.P. 68528, 21945-970, Rio de Janeiro, RJ, Brazil}
\author{Ehsan Khatami}
\affiliation{Department of Physics, San Jose State University, San Jose, CA 95616, USA}
\author{Shuxiang Yang}
\affiliation{Department of Physics and Astronomy, Louisiana State University, Baton Rouge, LA 70803, USA}
\author{Valery Rousseau}
\affiliation{Department of Physics and Astronomy, Louisiana State University, Baton Rouge, LA 70803, USA}
\author{Mark Jarrell}
\affiliation{Department of Physics and Astronomy, Louisiana State University, Baton Rouge, LA 70803, USA}
\author{Juana Moreno}
\affiliation{Department of Physics and Astronomy, Louisiana State University, Baton Rouge, LA 70803, USA}
\author{Randall G. Hulet}
\affiliation{Department of Physics and Astronomy and Rice Quantum Institute, Rice University, Houston, TX 77005, USA}
\author{Richard T. Scalettar}
\affiliation{Department of Physics, University of California, Davis, CA 95616, USA}

\maketitle

%%%%%%%%%%%%%%%%%%%%%%%%%%%%%%%%%%%%%%%%%%%%%%%%%%%%%%%%%%%%%%%%%%%%%%%%%%%%%%%%
% FIGURE 1
%%%%%%%%%%%%%%%%%%%%%%%%%%%%%%%%%%%%%%%%%%%%%%%%%%%%%%%%%%%%%%%%%%%%%%%%%%%%%%%%
\begin{figure}[b]
\includegraphics[width=1\columnwidth,height=2.2in]{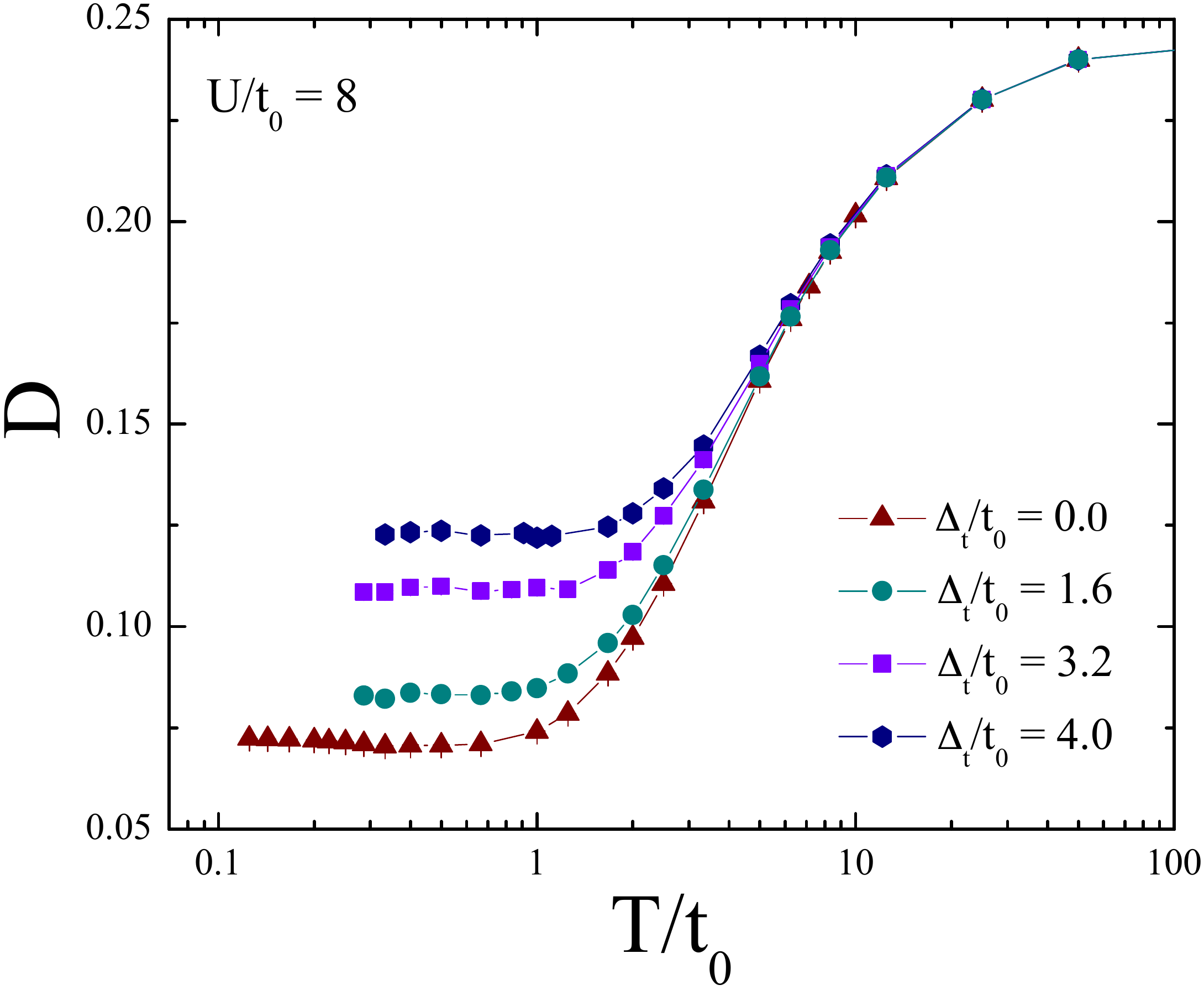}
\caption{
Double occupancy vs temperature for the clean and disordered systems with bond 
disorder strength $\Delta_t=0.0-4.0$. Unlike the case of chemical potential 
disorder, there is no qualitative change in the structure of $D(T)$:  double 
occupancy decreases at the temperature scale $T \sim U$ for all $\Delta_t$.
\label{fig:doccU8dt}}
\end{figure}

In this Supplement,
we present further quantum Monte Carlo (QMC) data which (1) analyzes $D(T)$ in case 
of bond disorder; (2) presents results at $U/t_0=4,12$ to expand upon the 
$U/t_0=8$ data in the main paper; (3) examines the shifts in the specific heat 
$C(T)$ curve, and its kinetic and potential energy components, as disorder is 
increased; (4) presents entropy $S$ as a function of disorder $\Delta_\mu$ at 
fixed temperature $T$; (5) connects to previous determinant QMC (DQMC) and dynamical cluster 
approximation (DCA) studies of the effects of disorder on magnetic transitions 
and conductivity, and experimental work in 1D; and (6) discusses more general 
disorder distributions.

%%%%%%%%%%%%%%%%%%%%%%%%%%%%%%%%%%%%%%%%%%%%%%%%%%%%%%%%%%%%%%%%%%%%%%%%%%%%%%%%
% FIGURE 2
%%%%%%%%%%%%%%%%%%%%%%%%%%%%%%%%%%%%%%%%%%%%%%%%%%%%%%%%%%%%%%%%%%%%%%%%%%%%%%%%
\begin{figure}[b]
\includegraphics[width=1.0\columnwidth,height=3.3in]{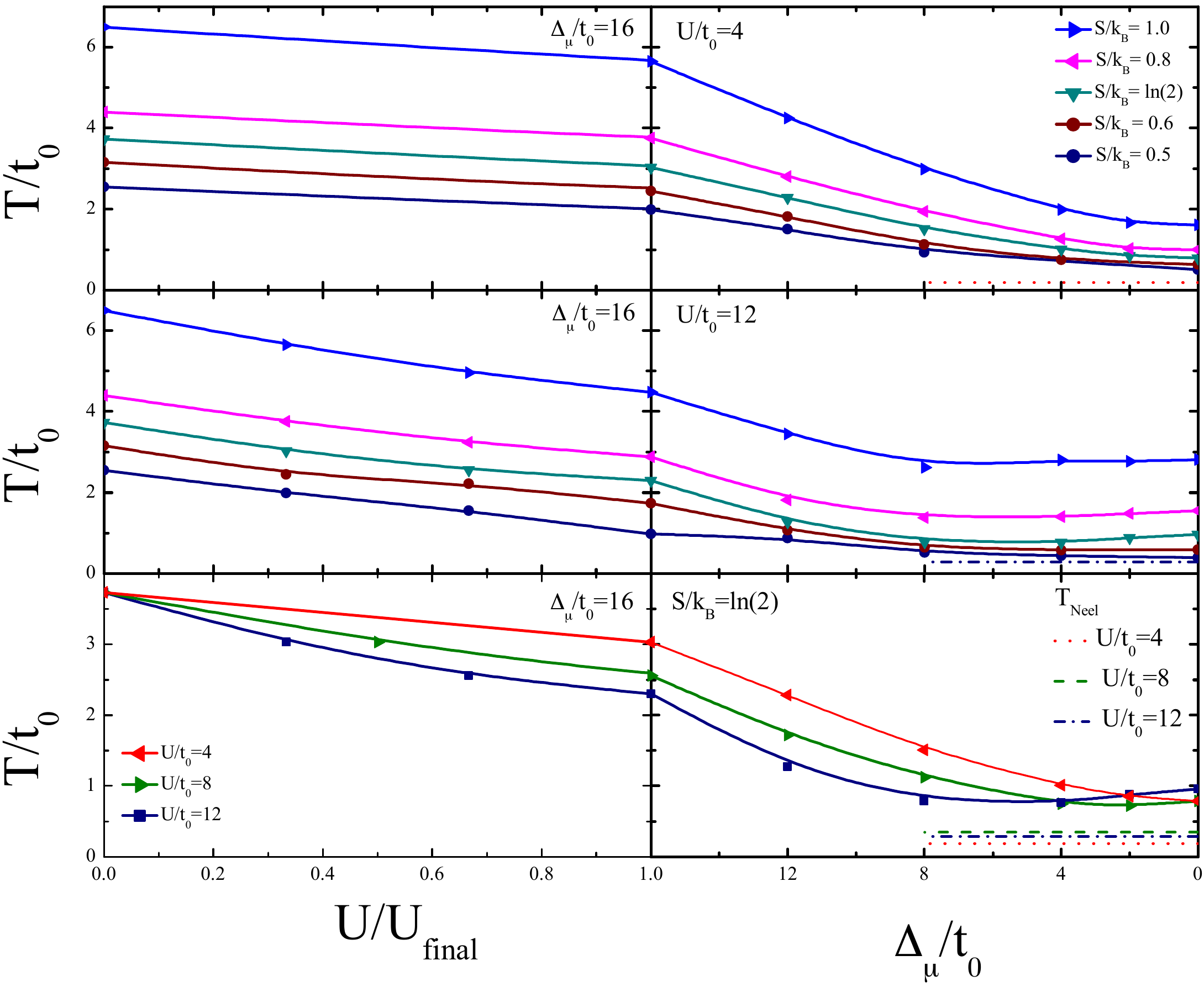}
\caption{
QMC results for the adiabats of the disordered 3D Hubbard model along paths which 
combine an increase of $U/t_0$ from 0 to 4 (top row, left) and to 12 (central row, left), both at fixed 
$\Delta_\mu/t_0=16$. This trajectory is followed by a reduction of the disorder 
to zero at constant $U$ (top and middle rows, right). These $U/t_0=4,12$ data 
complement the $U/t_0=8$ results in the main text. The bottom row compares the 
$S/k_B={\rm ln 2}$ adiabats for $U/t_0=4, 8, 12$. Note the horizontal axis of left panels
is $U/U_{\rm final}$ to allow results for different $U/t_0$ to be displayed together.
\label{fig:protocol3}}
\end{figure}

%%%%%%%%%%%%%%%%%%%%%%%%%%%%%%%%%%%%%%%%%%%%%%%%%%%%%%%%%%%%%%%%%%%%%%%%%%%%%%%%
% 1. DOUBLE OCCUPANC
%%%%%%%%%%%%%%%%%%%%%%%%%%%%%%%%%%%%%%%%%%%%%%%%%%%%%%%%%%%%%%%%%%%%%%%%%%%%%%%%
\vskip 0.2cm
\noindent
{\bf 1. Double occupancy.}

Figure~S\ref{fig:doccU8dt} shows the double occupancy $D(T)$ for hopping 
disorder.  The effect of randomness is very different from the case with 
chemical potential disorder: there is some enhancement of $D$, but the
basic structure remains unaltered from the clean limit.  $D(T)$ falls
with decreasing $T$ at a charge fluctuation scale associated with $U$,
reaches a minimum at $T/t_0\sim 1.0$, and then has a region where $dD/dT$ 
can be slightly negative at low temperatures.  The effect of $\Delta_t$ is
only to shift the values of $D$ upwards (from $D \sim 0.07$ to $D \sim
0.12$ at the largest $\Delta_t=4.0$), in an almost temperature
independent way at low $T$.  These data suggest that disorder-induced
cooling is not uniquely linked to a negative $dD/dT$, as is the case
with Pomeranchuk cooling.

%%%%%%%%%%%%%%%%%%%%%%%%%%%%%%%%%%%%%%%%%%%%%%%%%%%%%%%%%%%%%%%%%%%%%%%%%%%%%%%%
% 2. WEAK AND STRONG COUPLING
%%%%%%%%%%%%%%%%%%%%%%%%%%%%%%%%%%%%%%%%%%%%%%%%%%%%%%%%%%%%%%%%%%%%%%%%%%%%%%%%
\vskip0.20in
\noindent
{\bf 2. Weak and strong coupling.}

Figure 2 of the main text showed the evolution of temperature following a 
protocol starting with a trapped gas in a disordered potential $\Delta_\mu/
t_0=16$ and with no interaction $U/t_0=0$, through a ramp of the lattice to 
$U/t_0=8$, followed by turning off the disorder $\Delta_\mu=0$. 
Figure S\ref{fig:protocol3} presents similar data at weaker ($U/t_0=4$) and
stronger $(U/t_0=12$) couplings.

%%%%%%%%%%%%%%%%%%%%%%%%%%%%%%%%%%%%%%%%%%%%%%%%%%%%%%%%%%%%%%%%%%%%%%%%%%%%%%%%
% FIGURE 3
%%%%%%%%%%%%%%%%%%%%%%%%%%%%%%%%%%%%%%%%%%%%%%%%%%%%%%%%%%%%%%%%%%%%%%%%%%%%%%%%
\begin{figure}[t]
\includegraphics[width=1\columnwidth]{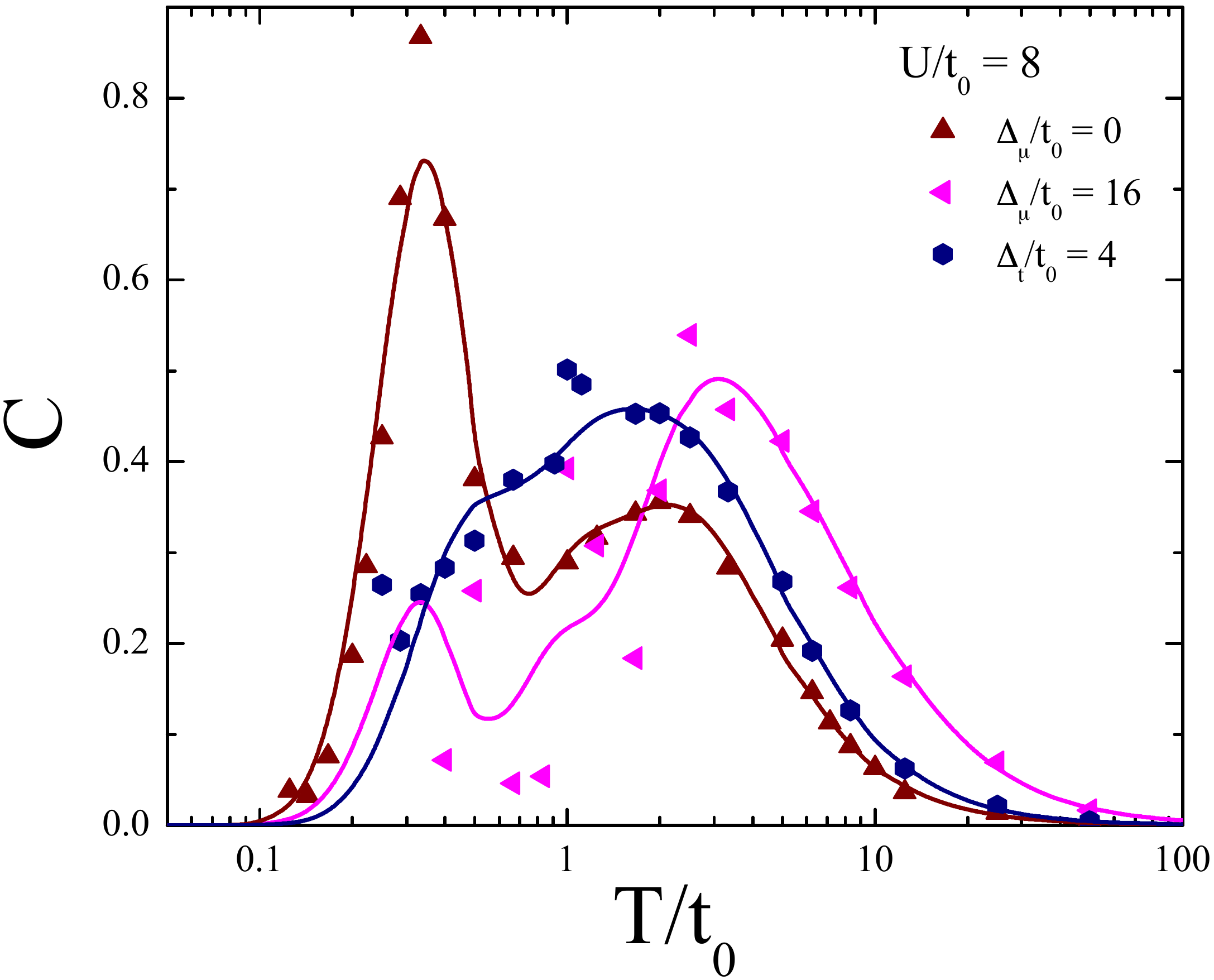}
\caption{
The specific heat $C(T)$ has two features in the clean limit: a high temperature 
($T \sim U$) peak associated with moment formation, and a low temperature peak 
$T \sim J \sim 4t^2/U$ where moment ordering occurs. As disorder increases, these 
peaks are smeared, and the low $T$ moment ordering peak especially is disrupted.  $C(T)$ structure resulting from excitations linked to
occupying sites with large local chemical potential appears at temperatures $T$ 
which scale with $\Delta_\mu$. Symbols represent the numerical 
derivative of the energy with respect to temperature, whereas lines were obtained 
from fitting the QMC energy data to a physically motivated function: 
$E(T)=E(0)+\sum_{l=1}^M c_l {\rm exp}(-  l \Delta/T)$ \cite{andy}. We have used 
$M=7$ or 8 exponentials to fit the data. 
\label{fig:CU8}}
\end{figure}

The main message is that disorder cooling occurs through all values of $U/t_0$, 
an important practical observation because a confining potential not only 
changes the density across the trap, but also the lattice depth. Figure
S\ref{fig:protocol3}, in combination with Fig.~2 of the main text, indicate some 
interesting trends with $U/t_0$. Since cooling takes place during the first
phase (left panels) of the protocol, when the interaction is increased in the 
presence of disorder, the larger $U/t_0$, the greater the cooling.  However, the 
cooling in the second phase (right panels), when the disorder is turned off, is 
less at larger $U/t_0$ so that the final temperature after the full protocol
is almost identical ($T/t_0 \sim 0.8$) for $U/t_0=4$ and $U/t_0=8$ when 
$S/k_B=\ln 2$, and is only slightly higher ($T/t_0 \sim 1$) for $U/t_0=12$.

%%%%%%%%%%%%%%%%%%%%%%%%%%%%%%%%%%%%%%%%%%%%%%%%%%%%%%%%%%%%%%%%%%%%%%%%%%%%%%%%
% 3. SPECIFIC HEAT
%%%%%%%%%%%%%%%%%%%%%%%%%%%%%%%%%%%%%%%%%%%%%%%%%%%%%%%%%%%%%%%%%%%%%%%%%%%%%%%%
\vskip 0.2cm
\noindent
{\bf 3. Specific Heat.}

The mechanism of entropy reduction at low $T$ is associated
with the modification and redistribution of the energy levels.
When $\Delta_\mu$ is sufficiently large, the weight under the specific
heat curve C(T), Fig.~S\ref{fig:CU8}, is shifted to higher temperatures.
This occurs both because of the reduction in magnetic ordering, and hence,
of the peak in $C(T)$ at $T \sim J$,  and also because of the introduction 
of new, high energy excitation modes associated with the occupation of sites 
with large $\mu_i$.  The shift in the structure of $C(T)$ to higher $T$ 
signals the reduction of $S(T)$ at low $T$, and therefore the phenomenon of 
disorder cooling.

\begin{figure}[t]
\includegraphics[width=1\columnwidth]{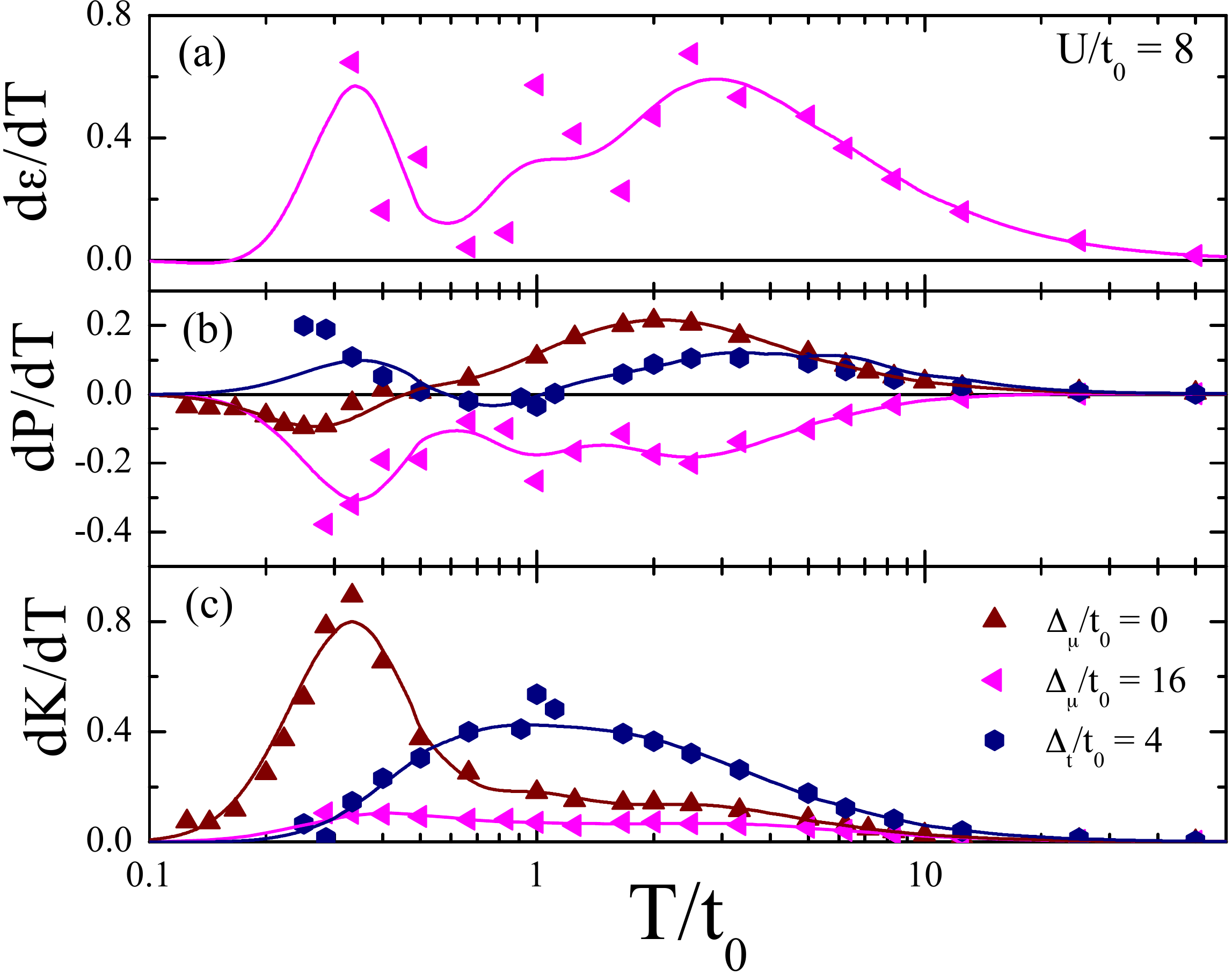}
\caption{
Temperature derivatives of the different components of the energy. The greatest 
qualitative changes are in the sign of $dP/dT$ for $\Delta_\mu/t_0=0$ versus
$\Delta_\mu/t_0=16$, and the reduction of the low $T$ peak in $dK/dT$.
For $\Delta_\mu/t_0=16$ lines were obtained from fitting the QMC data to 
$f(T)=f(0)+\sum_{l=1}^M c_l {\rm exp}(-  l \Delta/T)$ \cite{andy}.
\label{fig:KEU8}}
\end{figure}

Figure S\ref{fig:KEU8} shows the temperature derivatives of the kinetic energy
$K = 1/N \langle \sum_{\langle ij \rangle}
( c^{\dagger}_{i\sigma} c^{\phantom{\dagger}}_{j\sigma} +
c^{\dagger}_{j\sigma} c^{\phantom{\dagger}}_{i\sigma} ) \rangle $,
potential energy $P=U\times D$, %\langle n_\uparrow n_\downarrow \rangle$, 
and random chemical potential energy $ \epsilon = 
- 1/N \langle \sum_i \mu_i \, ( n_{i\uparrow} + n_{i\downarrow} ) \rangle$.
Together these derivatives sum to $C(T)$. The effect of disorder is most 
dramatic on $dP/dT$, which is mostly positive for $\Delta_\mu/t_0=0$ but 
negative for $\Delta_\mu/t_0=16$:  the double occupancy is shrinking as $T$ 
increases in the highly disordered limit.  This is also clearly seen in the 
inset of Fig.~1 of the main text. A second effect of disorder is to eliminate 
the low $T$ peak in $dK/dT$.  As site disorder grows, neighboring sites are 
no longer in resonance, and hopping is diminished.  This diminished hopping
is also linked to reduced local AF order.

%%%%%%%%%%%%%%%%%%%%%%%%%%%%%%%%%%%%%%%%%%%%%%%%%%%%%%%%%%%%%%%%%%%%%%%%%%%%%%%%
% FIGURE 4S
%%%%%%%%%%%%%%%%%%%%%%%%%%%%%%%%%%%%%%%%%%%%%%%%%%%%%%%%%%%%%%%%%%%%%%%%%%%%%%%%

%%%%%%%%%%%%%%%%%%%%%%%%%%%%%%%%%%%%%%%%%%%%%%%%%%%%%%%%%%%%%%%%%%%%%%%%%%%%%%%%
% 4. DEPENDENCE OF ENTROPY ON DISORDER
%%%%%%%%%%%%%%%%%%%%%%%%%%%%%%%%%%%%%%%%%%%%%%%%%%%%%%%%%%%%%%%%%%%%%%%%%%%%%%%%
\vskip 0.2cm
\noindent
{\bf 4. Dependence of entropy on disorder}

\begin{figure}[t]
\includegraphics[width=1\columnwidth]{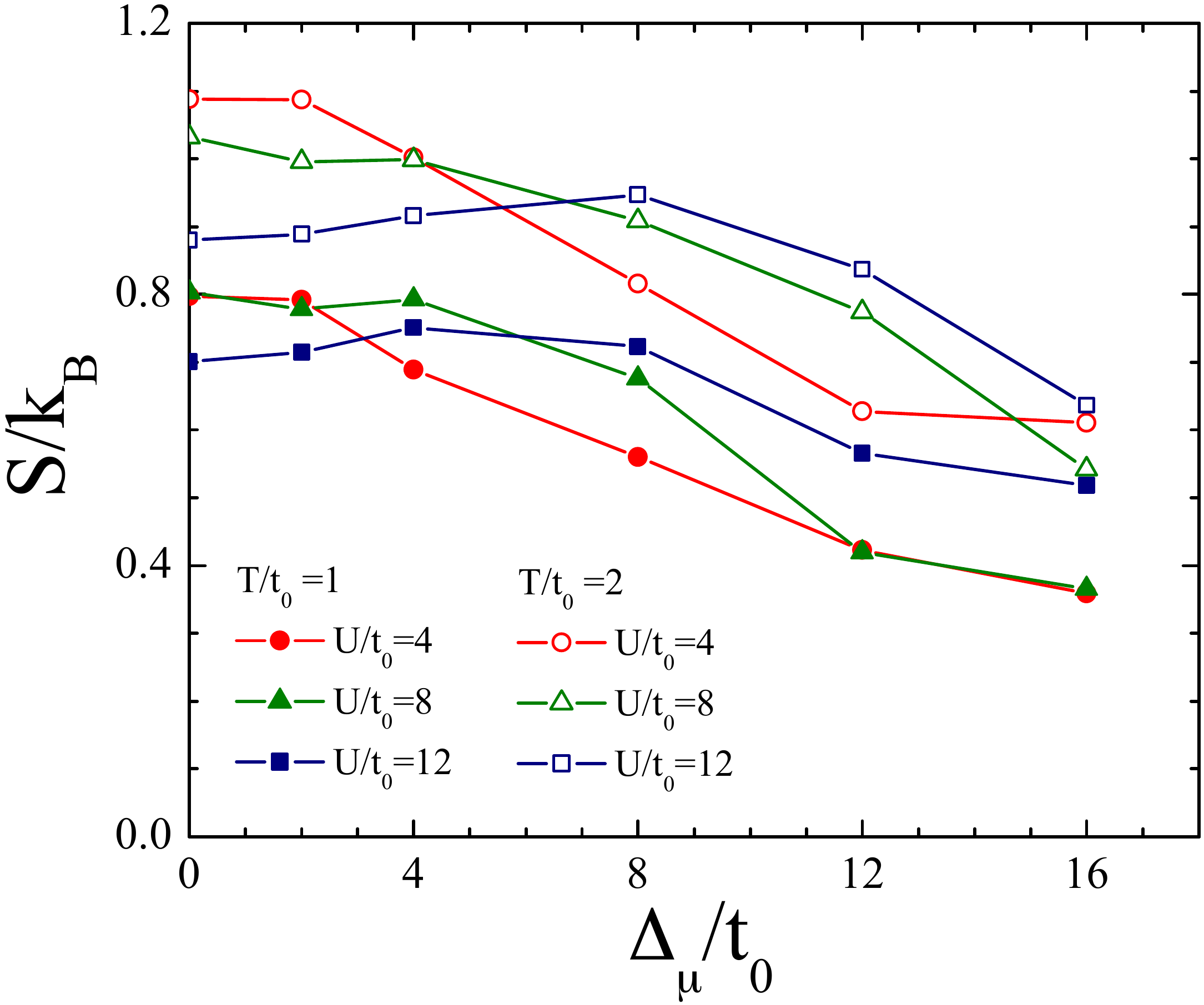}
\caption{
Entropy $S$ as a function of site disorder strength $\Delta_\mu$ for several 
fixed temperatures and interaction strengths. As $\Delta_\mu$ increases, the 
entropy is lowered.
\label{fig:ST1}
}
\end{figure}

Another way of looking at cooling is to examine the entropy as a function of 
$\Delta_\mu$ at fixed temperature $T$ (Fig.~S\ref{fig:ST1}).  This plot can be
inferred by taking a vertical cut  through the lines in Fig.~1 of the main text.
As $\Delta_\mu$ increases, the entropy is lowered.  Thus, if it is possible to 
cool an atomic gas to a given $T$ regardless of disorder strength, lower initial 
$S_i$ can be accessed.  If the disorder is then turned off adiabatically 
($S_f=S_i$), a lower final $T_f$ is achieved.  Possible approaches to cooling 
to the same $T_i$ in the presence of disorder, such as putting the disordered 
gas in thermal contact with a clean gas or a region of the trap with no disorder, 
were discussed in the Concluding Remarks of the main text.

%%%%%%%%%%%%%%%%%%%%%%%%%%%%%%%%%%%%%%%%%%%%%%%%%%%%%%%%%%%%%%%%%%%%%%%%%%%%%%%%
% FIGURE 5S
%%%%%%%%%%%%%%%%%%%%%%%%%%%%%%%%%%%%%%%%%%%%%%%%%%%%%%%%%%%%%%%%%%%%%%%%%%%%%%%%

%%%%%%%%%%%%%%%%%%%%%%%%%%%%%%%%%%%%%%%%%%%%%%%%%%%%%%%%%%%%%%%%%%%%%%%%%%%%%%%%
% 5. RELATED DQMC AND EXPERIMENTAL WORK
%%%%%%%%%%%%%%%%%%%%%%%%%%%%%%%%%%%%%%%%%%%%%%%%%%%%%%%%%%%%%%%%%%%%%%%%%%%%%%%%
\vskip0.20in
\noindent
{\bf 5. Related DQMC and experimental work.}

DQMC results for the effect of site and bond disorder on antiferromagnetism, 
analogous to Fig.~4 of the main text, have been reported in two dimensions
\cite{ulmke97,ulmke98} and in the limit of infinite dimensions
\cite{ulmke95,ulmke99}. In the former case, ordering occurs only at $T=0$.
An interesting effect in the latter case is the initial enhancement of 
$T_N$ when the chemical potential disorder is small. This occurs as 
a consequence of an increase in exchange energies.  For two neighboring sites
with site energy difference $\Delta_\mu$ the clean system exchange is altered 
to $J= 2t^2/(U - \Delta_\mu) + 2t^2/(U +\Delta_\mu) > 4t^2/U=J_{\rm clean}$
(See main text). The effect of disorder on the conductivity, and the possibility
of an Anderson insulator to metal transition, have also been studied
\cite{denteneer99,denteneer01,denteneer03}. An interesting feature of the 
transport properties of the disordered Hubbard Hamiltonian is the key role of 
particle-hole symmetry, i.e.~whether the disorder is in the site energies, which 
breaks particle-hole symmetry, or the bond hoppings, which preserves it.

The DCA is a powerful alternate approach to the thermodynamics of the 3D Hubbard
Hamiltonian~\cite{fuchs11}.  A recently proposed extension of this method, the 
typical medium DCA (TMDCA)~\cite{ekuma14} allows the treatment of disorder, 
generalizing earlier approaches~\cite{dobro03}, and accurately captures many 
features of the Anderson localization transition, including the critical 
disorder strength, the exponent of the typical density of states at the 
transition, and reentrant behavior. Given the usefulness of past comparisons 
of DQMC~\cite{paiva11} and DCA~\cite{fuchs11} in the absence of randomness, it 
will be useful to incorporate interactions in the TMDCA and study the disorder 
cooling proposed here.

A recent study of relaxation from an initial charge density wave state
in the {\it one dimensional} Hubbard Hamiltonian with a quasi-periodic
potential~\cite{schrieber15} also provides experimental and exact
diagonalization values for the critical disorder strength for MBL.
After adjusting for the difference in bandwidth, the size of
$\Delta_\mu$ we propose for disorder cooling lies close to this
threshold.  Given the strong tendency for localization in $d=1$, (much
more than a simple bandwidth renormalization) these results indicate the
$\Delta_\mu$ reported here are not in a regime where the system will
fail to equilibrate.  This work~\cite{schrieber15} 
also compares relaxation data with and without the presence of a trap
and shows they differ only in the vicinity of the MBL transition,
suggesting the inclusion of a trap in our 3D calculations will not
affect the qualitative results, except perhaps near $T_N$.

%%%%%%%%%%%%%%%%%%%%%%%%%%%%%%%%%%%%%%%%%%%%%%%%%%%%%%%%%%%%%%%%%%%%%%%%%%%%%%%%
% 6. MORE GENERAL DISORDER DISTRIBUTIONS
%%%%%%%%%%%%%%%%%%%%%%%%%%%%%%%%%%%%%%%%%%%%%%%%%%%%%%%%%%%%%%%%%%%%%%%%%%%%%%%%
\vskip 0.2cm
\noindent
{\bf 6. More general disorder distributions.}

There are a number of possible refinements of the calculations we have
reported here.  We have assumed a box distribution of the random
chemical potentials $\mu_i$ and near-neighbor hoppings $t_{ij}$. In
fact, the precise functional forms of these distributions from optical
speckle are known~\cite{speckle,white09,zhou10}.  Whereas we have studied 
site and bond randomness separately, the disorder-induced cooling produced
by hopping disorder tends to occur at lower $T$ than for chemical
potential disorder, making the possible interplay of the two of interest.